\documentstyle[twoside,fleqn,npb,epsfig]{article}
\newcommand{\AmS}{{\protect\the\textfont2
  A\kern-.1667em\lower.5ex\hbox{M}\kern-.125emS}}

\def\gg{\gamma\gamma}
\def\ppbar{\overline{\mbox p}\mbox{p}}

\def\sqee{\sqrt{s}_{\rm ee}}

\def\xg{x_{\gamma}}
\def\xgp{x_{\gamma}^+}
\def\xgm{x_{\gamma}^-}
\def\xgpm{x_{\gamma}^{\pm}}
\def\etajet{\eta^{\rm jet}}

\def\ee{\mbox{e}^+\mbox{e}^-}

\def\ETJET{E^{\rm jet}_T}
\def\pt{p_{\rm T}}
\def\dspt{{\rm d}\sigma/{\rm d}p_{\rm T}} 
\def\dset{{\rm d}\sigma/{\rm d}\ETJET}

\renewcommand{\thefootnote}{\fnsymbol{footnote}}

\title{Jet and hadron production in photon-photon collisions\footnotemark[1]}

\author{Stefan S\"oldner-Rembold \address{
        Universit\"at Freiburg, Hermann-Herder-Str.3, 
        D-79104 Freiburg im Breisgau, Germany}
        \thanks{Heisenberg-Fellow}}
\begin{document}

\begin{abstract}
Di-jet and inclusive charged
hadron production cross-sections measured in $\gg$ collisions by OPAL 
are compared to NLO pQCD calculations. Jet shapes measured in
$\gg$ scattering by OPAL, in deep-inelastic ep scattering
by H1 and in $\gamma$p scattering by ZEUS are shown to be consistent in similar
kinematic ranges. New results from TOPAZ on prompt photon
production in $\gg$ interactions are presented. 
\end{abstract}

\maketitle
\section{Leading Order parton processes}
The interaction of quasi-real photons ($Q^2\approx0$) studied at LEP and
the interaction of a quasi-real photon with a proton studied
at HERA (photoproduction)
are very similar processes. 
In leading order (LO) different event classes can be defined
in $\gg$ and $\gamma$p interactions. The photons can either
interact as `bare' photons (``direct'') 
\begin{figure}[h]
\begin{center}
\epsfig{file=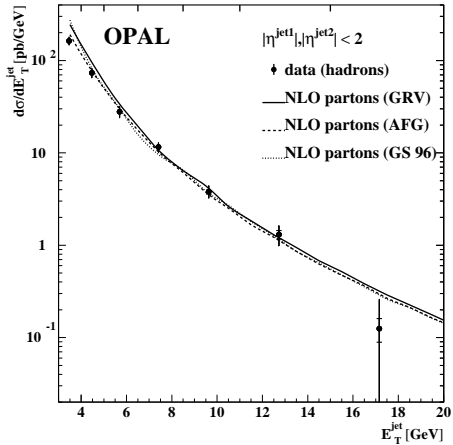,
width=0.37\textwidth}
\end{center}
\begin{center}
\caption{\label{fig-ettwojet}
The inclusive $\ee$ di-jet cross-section as a function
of $\ETJET$ for jets with $|\etajet|<2$.}
\end{center}
\end{figure}
or as hadronic fluctuations (``resolved''). 
Direct and resolved interactions can be separated by measuring
the fraction $\xg$ of the photon's momentum participating in the 
hard interaction for the two photons. In $\gg$ interactions
they are labelled $\xgpm$ for the two photons.
\setcounter{footnote}{0}
\renewcommand{\thefootnote}{\fnsymbol{footnote}}
\footnotetext[1]{
submitted to the proceedings of DIS99, DESY-Zeuthen, Berlin, Germany,
April 1999 
}

Ideally, the direct $\gg$ events with two bare
photons are expected to have $\xgp=1$ and
$\xgm=1$,
whereas for double-resolved events both values $\xgp$ and $\xgm$
are expected to be much smaller than one. 
In photoproduction, the
interaction of a bare photon with the proton is labelled `direct'
(corresponding to `single-resolved' in $\gg$) and
the interaction of a hadronic photon is called `resolved'
(corresponding to `double-resolved' in $\gg$).

\section{Di-jet production}
\label{sec-jet}
Studying jets gives access to the parton dynamics
of $\gg$ interactions.
OPAL has therefore measured di-jet production in $\gg$ scattering
at $\sqee=161-172$~GeV using the cone jet finding algorithm with 
$R=1$~\cite{bib-opaljet}. 

The differential cross-section $\dset$
for di-jet events with pseudorapidities 
$|\etajet|<2$ 
is shown in Fig.~\ref{fig-ettwojet}. The measurements are
compared to a parton level NLO calculation~\cite{bib-klasen} 
for three different NLO parametrisations of the parton distributions
of the photon
GRV-HO~\cite{bib-grv}, AFG~\cite{bib-afg} and GS~\cite{bib-gs}.
The calculations using the three different
NLO parametrisations are in good agreement with
the data points except in the first bin where
theoretical and experimental uncertainties are large.

\begin{figure}[htbp]
\begin{center}
\epsfig{file=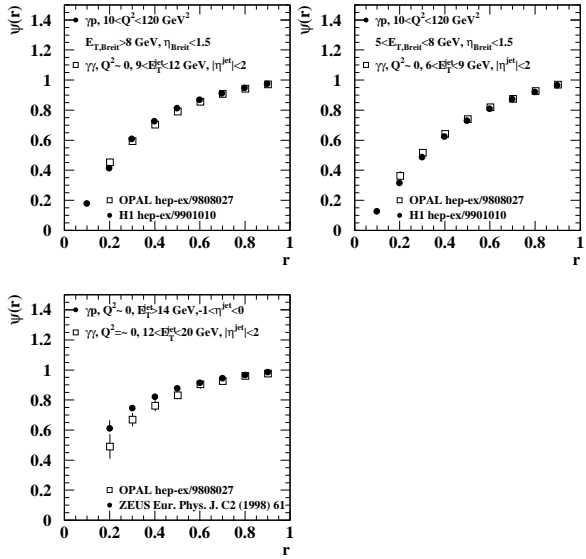,
width=0.48\textwidth}
\end{center}
\label{fig-psi}
\caption{A comparison of jet shapes $\psi(r)$ measured in
di-jet events by OPAL, H1 and ZEUS.}
\end{figure}
\section{Jet shapes}
\label{sec-shape}
The jet shape is characterised by the 
fraction of a jet's transverse energy ($\ETJET$) that lies inside
an inner cone of radius $r$ concentric with the jet defining
cone:
\begin{equation}
  \psi(r) = \frac{1}{N_{\rm jet}} \sum_{\rm jets} 
\frac{E_{\rm T}(r)}{E_{\rm T}(r=R)},
\end{equation}
where $E_{\rm T}(r)$ 
is the transverse energy within the inner cone of radius $r$
and $N_{\rm jet}$ is the total number of jets in the sample. 
The jet shapes are corrected to the hadron level using the
Monte Carlo. It has been shown by OPAL that the jets become narrower
with increasing $\ETJET$, that the jet shapes are nearly independent
of $\etajet$ and 
that gluon jets are broader than quark jets~\cite{bib-opaljet}.

The measured jet shapes are compared
to data from the HERA experiments
in similar kinematic ranges. H1 has measured jet shapes
for di-jet events produced in deep-inelastic scattering (DIS) 
with $10<Q^2<120$~GeV$^2$ and 
$2\cdot10^{-4}<x<8\cdot10^{-3}$~\cite{bib-h1shape}.
The events are boosted into the Breit-frame. 
ZEUS has mesured jet shapes in di-jet photoproduction
for quasi-real photons in the lab frame~\cite{bib-zeusshape}.
In the regions $\eta_{\rm Breit}<1.5$ (H1 data) and
and $-1<\etajet<0$ (ZEUS data), most jets should be quark jets.
The same is true for the OPAL data at relatively large $\ETJET$,
where the direct process dominates. A comparison shown in
Fig.~\ref{fig-psi} performed for similar transverse energy ranges
shows good agreement of the OPAL $\gg$ and the H1 DIS data.
The jets in photoproduction events measured by ZEUS 
are narrower than the $\gg$ jets. This could be due
to the slightly larger $\ETJET$ in the ZEUS data.
A detailed comparison of jet widths measured in
$\gg$ interactions and other processes ($\ee,\ppbar,\gamma$p)
has recently been published by TOPAZ~\cite{bib-tjets}.

\section{Charged hadron production}
Hadron production at large transverse momenta
is also sensitive to the partonic structure of the interactions
without the theoretical and experimental problem related
to the various jet algorithms. Interesting comparisons
of $\gg$ and $\gamma$p data taken at LEP and HERA, respectively,
should be possible in the future, since similar hadronic
centre-of-mass energies $W$ of the order 100 GeV are accessible
for both type of experiments.

\begin{figure}[htbp]
   \begin{center}
      \mbox{
          \epsfxsize=0.36\textwidth
          \epsffile{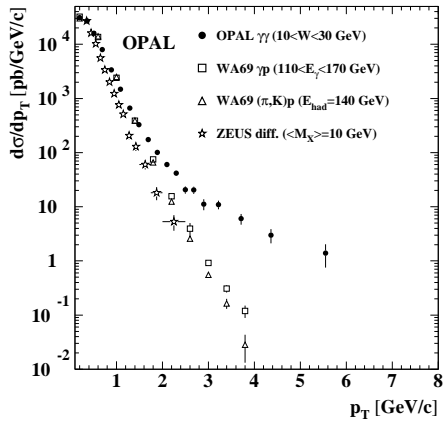}
          }
   \end{center}
\caption{\label{fig-wa69}
The $\pt$ distribution measured in $\gg$ interactions
in the range $10<W<30$~GeV is compared to the $p_{\rm T}$ distribution
measured in $\gamma$p and hp (h$=\pi,$K) interactions by 
WA69~\protect\cite{bib-wa69}. 
The cross-section values on the ordinate
are only valid for the OPAL data.}
\end{figure}

The distributions of the transverse momentum $\pt$ 
of hadrons produced in $\gg$ interactions 
are expected to be harder than in $\gamma$p or hadron-p
interactions due to the direct component. 
This is demonstrated in Fig.~\ref{fig-wa69} by comparing 
$\dspt$ for charged hadrons measured in $\gg$ interactions by 
OPAL~\cite{bib-opalhad} to the $p_{\rm T}$ distribution 
measured in $\gamma$p and hp (h$=\pi,$K) interactions by WA69~\cite{bib-wa69}. 
The WA69 data are normalised to the $\gg$ data in the low 
$p_{\rm T}$ region 
at $\pt\approx 200$~MeV/$c$ using the same factor for the hp and the
$\gamma$p data.
The hadronic invariant
mass of the WA69 data is about $W=16$~GeV which is
of similar size as the average $\langle W \rangle$ of the $\gg$
data in the range $10<W<30$~GeV.

Whereas only a small increase is observed
in the $\gamma$p data compared to the h$\pi$ data at large $\pt$,
there is a significant increase of the relative rate in the range 
$\pt>2$~GeV/$c$ for $\gg$ interactions due to the
direct process. 

The $\gg$ data are also compared to
a ZEUS measurement of charged particle production 
in $\gamma$p
events with a diffractively dissociated photon at $\langle W \rangle = 
180$~GeV~\cite{bib-zeus}
The invariant mass relevant for this comparison 
should be the mass $M_{\rm X}$ of the
dissociated system (the invariant mass of the `$\gamma$-Pomeron' system).
The average $\langle M_{\rm X} \rangle$ equals 10 GeV for the data shown.
The $\pt$ distribution falls exponentially, similar to the
$\gamma$p and hadron-p data, and shows no flatening
at high $\pt$ due to a possible hard component of the Pomeron.

\begin{figure}[htbp]
   \begin{center}
      \mbox{
          \epsfxsize=0.3\textwidth
\epsffile{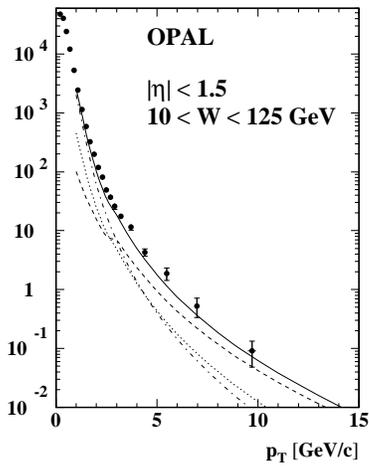}
          }
   \end{center}
\caption{\label{fig-dspt}
$\dspt$ for pseudorapidities $|\eta|<1.5$ in the range 
$10<W<125$~GeV compared to NLO calculations for $\pt>1$~GeV$/c$ 
(continuous curve) together with the double-resolved 
(dot-dashed), single-resolved (dotted)
and direct contributions (dashed).}
\end{figure}

NLO calculations~\cite{bib-binnewies} of the 
cross-sections $\dspt$ are shown in Fig.~\ref{fig-dspt}.
The cross-sections
are calculated using the QCD partonic cross-sections,
the NLO GRV parametrisation
of the parton distribution functions~\cite{bib-grv} and
fragmentation functions fitted to e$^+$e$^-$ data.
The renormalisation and factorisation scales
are set equal to $\pt$. 
The change in slope around $\pt=3$~GeV/$c$ in the
NLO calculation is due to the charm threshold.
The agreement between the data and the NLO
calculation is good. 

\section{Prompt photons}
The production of prompt photons in $\gg$ interactions
can also be used to measure the quark and gluon content of
the photon~\cite{bib-pp}. At TRISTAN energies the single-resolved
process
$\gamma \mbox{q}\to\gamma \mbox{q}$ is expected to dominate
the prompt photon production cross-section, whereas at LEP2
energies double-resolved processes should become important.
\begin{figure}[htbp]
\begin{center}
\begin{tabular}{cc}
\epsfig{file=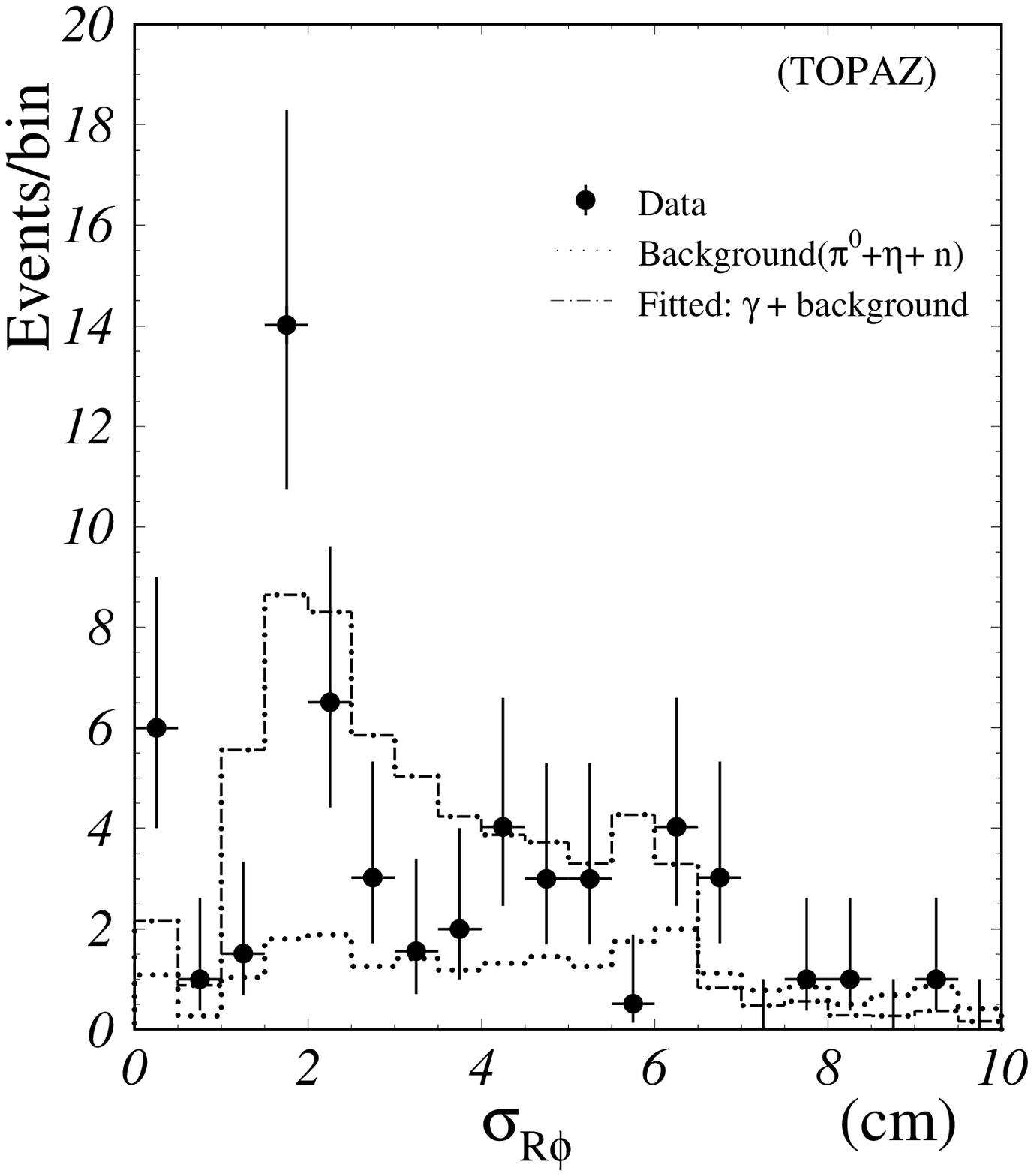,width=0.22\textwidth} &
\epsfig{file=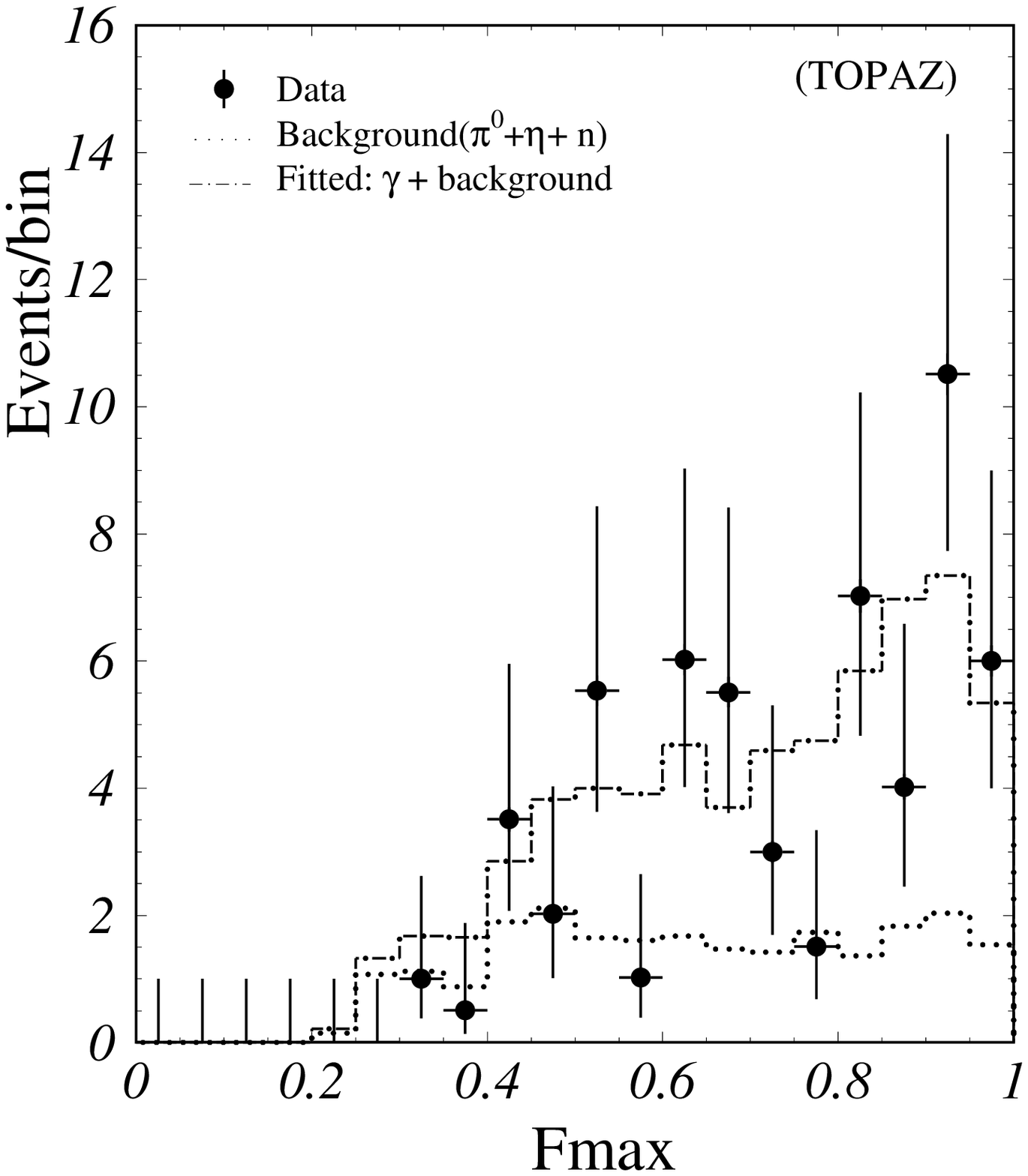,width=0.22\textwidth}
\end{tabular}
\end{center}
\label{fig-topaz}
\caption{Shower shape
parameters, $\sigma_{\rm R\phi}$ and $F_{\rm max}$, 
for the prompt photon events
measured by TOPAZ.}
\end{figure}

TOPAZ has measured the prompt photon cross-section
$\sigma(\ee\to\ee\gamma X)$ by fitting signal plus
background (mainly from $\pi^0$ decays) to variables describing the 
shower shape in the calorimeter (Fig.~\ref{fig-topaz}).
The variables are the rms of the cluster width
in the r$\phi$ direction, $\sigma_{\rm R\phi}$, and the
ratio of the maximum energy in a cluster to the total
cluster energy, $F_{\rm max}$.
 
TOPAZ obtains $\sigma(\ee\to\ee\gamma X)=(1.48 \pm 0.4 \pm 0.49)$~pb
for photons with energies greater than 2 GeV using
a data set with an integrated luminosity of $L=288$~pb. 
This result is about 1.5-2 standard deviations larger than
the LO cross-sections $\sigma(\ee\to\ee\gamma X)$ of $0.35$~pb and
$0.50$~pb which were obtained with PYTHIA using SaS-1D~\cite{bib-sas} and
LAC1~\cite{bib-lac},
respectively.  

\section*{Acknowledgements}
I want to thank Hisaki Hayashii for helping me
with the TOPAZ data, Michael Klasen for providing
the NLO calculations of the di-jet cross-sections,
Tancredi Carli for discussion on the jet shapes
and the organizers for this interesting and
enjoyable workshop.

\end{document}